\documentclass[jap,twocolumn,showpacs]{revtex4}
\usepackage{graphicx}
\usepackage{bm}
\begin{document}
\title{Chemical potential shift and gap-state formation in SrTiO$_{3-\delta}$ revealed by photoemission spectroscopy}
\author{Prabir Pal}
\email{palp@nplindia.org, Tel: +91-11-45609329}
\author{Pramod Kumar}
\author{Aswin V.}
\author{Anjana Dogra}
\author{Amish G. Joshi}
\affiliation{CSIR-National Physical Laboratory, Dr. K. S. Krishnan Road, New Delhi - 110012, India}

\begin{abstract}
In this study, we report on investigations of the electronic structure of SrTiO$_3$ annealed at temperature ranging between 550 and 840$^\circ$C in an ultrahigh vacuum. Annealing induced oxygen vacancies (O$_{vac}$) impart considerable changes in the electronic structure of SrTiO$_3$. Using core-level photoemission spectroscopy, we have studied the chemical potential shift ($\Delta\mu$) as a function of annealing temperature. The result shows that the chemical potential monotonously increases with electron doping in SrTiO$_{3-\delta}$. The monotonous increase of the chemical potential rules out the existence of electronic phase separation in the sample. Using valence band photoemission, we have demonstrated the formation of a low density of states at the near Fermi level electronic spectrum of SrTiO$_{3-\delta}$. The gap-states were observed by spectral weight transfer over a large energy scale of the stoichiometric band gap of SrTiO$_3$ system leading finally to an insulator - metal transition. We have interpreted our results from the point of structural distortions induced by oxygen vacancies.
\end{abstract}

\emph{Journal of Applied Physics {\bf 116}, 053704 (2014)}

\maketitle
\section{INTRODUCTION}

Stoichiometric SrTiO$_3$ is a perovskite-type band insulator with Ti$^{4+}$ compositions having an indirect band gap of 3.25 eV \cite{cardona}. It is usually used as a substrate for epitaxial growth of perovskite-type compound films. The heterostructures integrated with SrTiO$_3$ and other oxides have been found to exhibit ferromagnetism, superconductivity and insulator-metal (IM) transitions at the interface \cite{ariando,li,bert,dikin}. More recently, a lot of attention have been focused on transport and magnetic behavior of SrTiO$_{3-\delta}$ single crystals with oxygen vacancies (O$_{vac}$) due to their potential for designing functional materials with desired conductivity \cite{liu1,liu2}. The IM transitions can be found in electron doped SrTiO$_3$ surfaces where doping can be achieved by partial substitution of Sr (or Ti) by La (or Nb) \cite{fujimori01,ishida} or by creating O$_{vac}$ through ultrahigh vacuum (UHV) annealing \cite{liu1}. The IM transitions in electron doped SrTiO$_3$ are found to be closely related to the unique electronic structure derived from the Ti $3d$ and O $2p$ hybridized orbitals. The electronic behavior of doped compounds with Ti$^{4+}$ and Ti$^{3+}$ mixed valence systems is dominated mainly by the competition among the electron-phonon coupling, the electronic repulsion, and the kinetic energy of the carriers \cite{waser,gervais,van,takizawa,chang}. The near Fermi level (E$_F$) electronic behavior of these mixed valence compounds with O$_{vac}$ depend on the topology of the TiO$_6$ octahedra in their structure. The electron-electron interaction and the kinetic energy of the charge carriers are sensitive to the deformation of the TiO$_6$ octahedra with O$_{vac}$ in the lattice. There have been many photoemission studies highlighting the gap-states in the near E$_F$ region in La-doped and Nb doped SrTiO$_3$ thin films with no significant O$_{vac}$ \cite{fujimori01,ishida}. The density of states (DOS) in electron doped SrTiO$_3$ cannot be explained within the rigid band model, suggesting that the gap states observed in La doped and Nb doped SrTiO$_3$ may indeed have the same origin. However, the role of O$_{vac}$ in the electronic structure of SrTiO$_{3-\delta}$ has not yet been explored. By creating O$_{vac}$ through annealing, electrons are doped, which in general increases the chemical potential. The chemical potential is one of the most fundamental physical quantities in the electron doped systems and can be measured through the shifts of photoemission spectra. Recently, electronic phase separation is observed in metallic LaAlO$_3$$\setminus$SrTiO$_3$ interface \cite{ariando}. However, it is unclear whether the O$_{vac}$ in SrTiO$_3$ surface would also lead to such electronic phase separation. Such phase separation phenomena have attracted our attention because it results in the pinning of the chemical potential by the vacancy induced states formed within the band gap of stoichiometric SrTiO$_3$. Therefore, it is important to investigate possibility of such chemical potential pinning in the SrTiO$_{3-\delta}$ through the measuremnets of chemical potential shift when it is annealed under UHV conditions. Apart from charges, the orbital degrees of freedom also play an important role in this scenario \cite{yin}. The charge carriers are strongly influenced by the symmetry of the Ti $3d$ orbitals hybridized with the O $2p$ orbitals of the TiO$_6$ structures.

In order to understand the nature of the near E$_F$ spectral weight behavior in SrTiO$_{3-\delta}$ surface, we have studied the electronic structure changes of SrTiO$_3$ as a function of O$_{vac}$. The O$_{vac}$ created by UHV annealing can produce electrons, which will occupy the two neighboring Ti sites of the TiO$_2$ plane \cite{fang}. Thus, the vacancies are ionized and the Ti$^{4+}$ in the SrTiO$_3$ is reduced to Ti$^{3+}$. We studied the Ti$^{3+}$ evolution as a function of annealing temperature. We have deduced the chemical potential shift from the core level shifts of photoemission (PES) spectra as a function of annealing temperature. By creating O$_{vac}$ through annealing, electrons are doped, which raise the chemical potential without indication of pinning of chemical potential. The results show that there is no intrinsic electronic phase separation of charge carriers when SrTiO$_3$ is annealed up to 840$^\circ$C. The O$_{vac}$ concentration is proportional to the annealing temperature. With low filling of the Ti $3d$ band, these materials lead to the formation of new spectral features of Ti $3d$ like character below the Fermi level (E$_F$). Our studies show a shift in the near E$_F$ gap-states as a function of annealing. At sufficiently low O$_{vac}$ in the temperature range 550$^\circ$C - 725$^\circ$C the gap-states are localized at $\sim$ 3.0 eV below E$_F$ in the sense that the electrons are being trapped by the O$_{vac}$ and will not contribute to the conductivity of the sample. On the other hand, above 725$^\circ$C annealing temperature the gap-states become delocalized at $\sim$ 1.0 eV below E$_F$ and are responsible for the conductivity of the sample. The shifts of the gap-states with the IM transitions are highly non-rigid band like in the sense that the gap-states build up intensity at $\sim$ 1.0 eV below E$_F$ with increasing O$_{vac}$. 

\section{EXPERIMENT}

High purity single crystalline samples of SrTiO$_3$ (100) (Crystal-GmbH) were used in this study. The stoichiometric SrTiO$_3$ samples were heated in situ to different temperatures in the range 200$^\circ$C - 840$^\circ$C under UHV conditions and kept at constant temperature for more than 6 h, after which the samples were cooled down to room temperature before photoemission measurements. In all the cases, the base pressure inside the chamber during the thermal treatment was better than 1.0 x 10$^{-9}$ Torr. The cleaned surfaces were found from 550$^\circ$C annealing temperature onwards, as confirmed by the absence of carbon and surface oxygen peaks in PES spectra. We do not observe any time-dependent core-level energy shifts from 550$^\circ$C sample onwards measured within a day.

Photoemission measurements were performed using Omicron mu-metal UHV system equipped with a twin anode Mg/Al X-ray source (DAR400), a monochromator and a hemispherical electron energy analyzer (EA 125 HR). All the photoemission measurements were performed inside the analysis chamber under base vacuum of $\sim$ 1.0 x 10$^{-10}$ Torr using monochoromatized Al K$\alpha$ line with photon energy 1486.60 eV. The spectra reported here were obtained at an emission angle of 30$^\circ$. The binding energy calibration was done by measuring $Au$ 4f$_{7/2}$ and the Fermi energy (E$_F$) of an $Au$ film in electrical contact with the sample. The total energy resolution, estimated from the width of the Fermi edge, was about $400$ meV for monochomatic Al K$\alpha$ line with photon energy 1486.60 eV. The sample temperature was measured by using a calibrated thermocouple sensor touching the sample plate. Each temperature measurements including the reference sample data were completed within a day.

\begin{widetext}
\begin{table}[h]
\begin{center}
\caption{The fitting parameters for the Ti $2p$ core-level spectra are shown in Table 1. We have estimated the Ti$^{3+}$ component from the normalized area. Uncertainty in determining the values of atomic compositions is estimated to be $\pm$ 5$\%$.}
\vskip0.5cm
\begin{tabular}{|c|c|c|c|c|c|c|}
\hline
Annealing & Loss of & Ti$^{3+}$ & \multicolumn{4}{c|} {Ti $2p$} \\
\cline{4-7}
temperature & oxygen atoms& content & \multicolumn{2}{c|}{Ti $2p_{3/2}$}&\multicolumn{2}{c|}{Ti $2p_{1/2}$}\\
\cline{4-7}
($^\circ$C) & ($\delta$) & & Ti$^{3+}$ Position & Ti$^{4+}$ Position & Ti$^{3+}$ Position & Ti$^{4+}$ Position \\
&&&(eV)&(eV)&(eV)&(eV) \\
\cline{1-7}
550&0.004&0.013&457.567&459.564&463.330&465.282\\
\hline
600&0.007&0.020&457.564&459.565&463.340&465.281\\
\hline
660&0.012&0.032&457.622&459.625&463.360&465.362\\
\hline
725&0.011&0.030&457.649&459.654&463.390&465.392\\
\hline
780&0.013&0.035&457.627&459.625&463.370&465.373\\
\hline
840&0.019&0.052&457.716&459.715&463.440&465.452\\
\hline
\end{tabular}
\end{center}
\end{table}
\end{widetext}

\section{RESULTS AND DISCUSSION}

Figure 1 shows the core level photoemission spectra of the O $1s$, Ti $2p_{3/2}$ and Sr $3d$ recorded using monochoromatized Al K$\alpha$ line with photon energy $1486.60$ eV. Comparing the background to signal ratio there is a reduction of O $1s$ and Ti$^{4+}$ peak intensity with increasing annealing temperatures while Ti$^{3+}$ intensity increases with increasing annealing temperatures. On the other hand, the intensity of Sr $3d$ shows a marginal change with annealing temperatures. It is important to stress that in the temperature range 200$^\circ$C - 500$^\circ$C the amount of carbon and a satellite structure on the high binding energy side of the O $1s$ core level are progressively decreasing. The satellite structure on the high binding energy side of the O $1s$ core level is a signature of the oxygen contamination on the surface \cite{nagarkar}. For a quantitative estimate of the annealing dependent atomic ratio changes from the integral peak areas of the O $1s$, Ti $2p$ and Sr $3d$ core level spectra we have fitted the peaks using Lorentzian-Gaussian line shapes after background correction. The peak areas are then normalized to the inelastic mean free path, photoionization cross-section and analyzer transmission function (1st approximation). The nominal compositions of the different samples obtained from this procedure are listed in Table 1. The reduction of oxygen was $\sim$ 0.4, 0.7, 1.2, 1.1, 1.3 and 1.9$\%$ respectively in 550, 600, 660, 725, 780 and 840$^\circ$C samples. This confirms that the presence of O$_{vac}$ keep on increasing with an increase of annealing temperature. The weak Ti$^{3+}$ peak has $\sim$ $2$ eV lower binding energy than the Ti$^{4+}$ peak centered at $\sim$ $459.5$ eV binding energy which is consistent with other studies \cite{ishida,marshall1}. The spectra shown in Fig. 1(b), which were taken at different annealing conditions, demonstrate the transfer of spectral weight from Ti$^{4+}$ peak to Ti$^{3+}$ peak with UHV annealing. This corresponds to a mixed Ti$^{3+}$ and Ti$^{4+}$ state in SrTiO$_3$ resulting from the elimination of oxygen. The transfer of spectral weight is found to increase with increase in annealing temperatures. Thus we can infer that the samples which are annealed at high temperature under UHV conditions are more conductive than the low temperature annealed samples. The formation of O$_{vac}$ in SrTiO$_3$ (100) surface is proportional to the UHV annealing temperature.  

\begin{figure}[t]
\vskip 1.0cm
\includegraphics[width=4.0in]{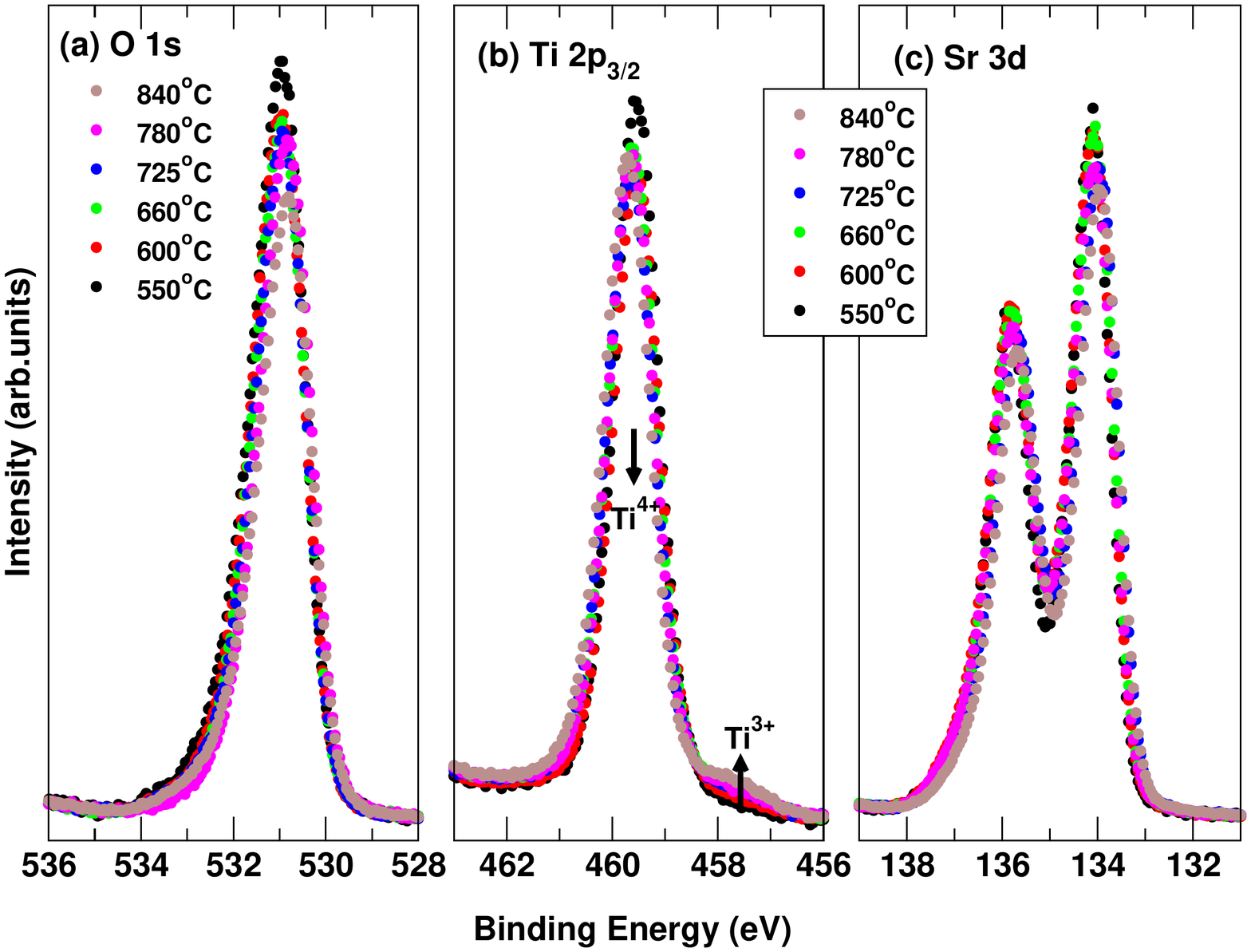}
\caption{\label{figure_1} (a) O $1s$ (b) Ti $2p_{3/2}$ and (c) Sr $3d$ core level photoemission spectra recorded using $1486.60$ eV photon energy. The spectra showed a progressive reduction of O $1s$ and Ti$^{4+}$ signal and an increase of Ti$^{3+}$ signal with increasing annealing temperature. While Sr $3d$ peak intensity does not change with annealing.} 
\end{figure}

The Ti $2p$ core level spectrum carries a substantial amount of physics involved in the properties of this material with annealing. For a quantitative estimate of the Ti$^{3+}$ content we have fitted the Ti $2p$ core-level peak. A Shirley background has been subtracted from the raw data. Results of the data fit are shown in Table 1 and Fig. 2. The spectra consist of two pair of lines and in each pair one appears as a shoulder on the low energy side of the other. The development of low energy side of the Ti $2p$ core level provides clear evidence for the presence of Ti$^{3+}$ ions and thus the formation of an electron doped state in the sample. The Ti$^{3+}$ component was $\sim$ 1$\%$ for 550 $^\circ$C sample and $\sim$ 5$\%$ for 840 $^\circ$C sample. The appearance of the Ti$^{3+}$ ions in the Ti $2p$ spectra is consistent with the observations of gap states in the valence band spectra (Fig. 5).

\begin{figure}[t]
\vskip 1.0cm
\includegraphics[width=4.0in]{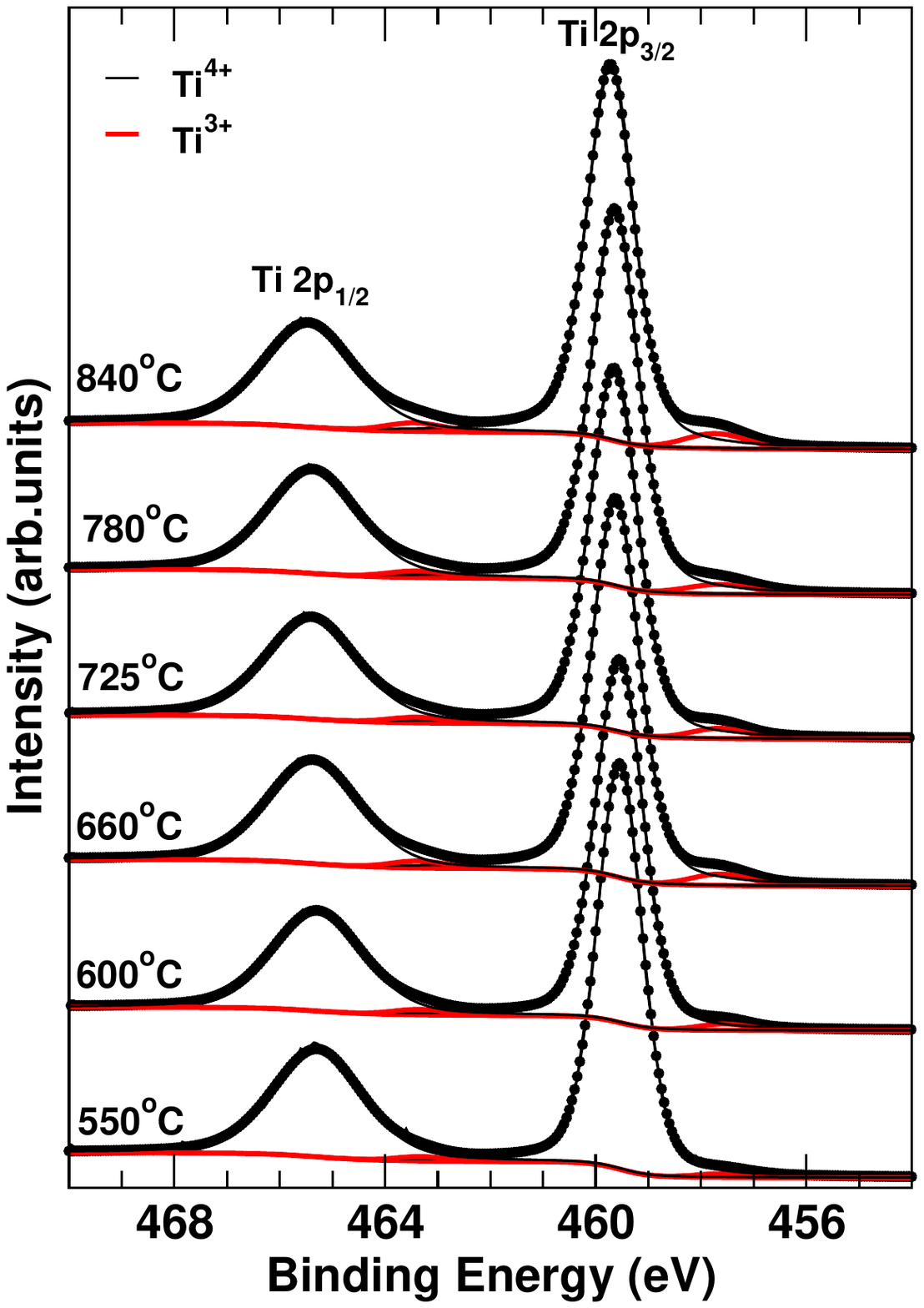}
\caption{\label{figure_2} Ti $2p$ core level photoemission spectra are fitted by four peaks. The Ti$^{3+}$ states are located at the low energy side of the Ti $2p$ core level. Ti$^{3+}$ and Ti$^{4+}$ peaks are shown in red and black respectively. We have used a Shirley background, which was kept the same for all the spectra.}
\end{figure}

Figure 3 shows the core level spectra shifted along the y axis by a constant for clarity. The vertical line demonstrates the estimated peak positions of the core levels. We have used the midpoint for the core levels since the line shape of the core levels does not change much with annealing. Here, we concentrate on the subtle changes in the binding energy of the core levels. The O $1s$ and Sr $3d$ core levels are shifted to lower binding energy while Ti $2p$ is shifted to higher binding energy with increasing annealing temperature. The shifts of Ti $2p$ core level to higher binding energy side are consistent with the earlier reports on n-doped SrTiO$_3$ using photoemission spectroscopy \cite{higuchi}. This shift in the core level spectra can be explained by the chemical potential shift. In order to estimate the chemical potential shift from a set of core-level spectra, we have used the formula that the shift of the binding energy ($\Delta$E$_B$) is given by $\Delta$E$_B$ = $\Delta\mu$ + $K$$\Delta Q$ + $\Delta$V$_M$ + $\Delta$E$_R$ \cite{hufner}, where $\Delta\mu$ is the change in the chemical potential, $K$ is the coupling constant of the coulomb interaction between the valence and core electrons, $\Delta Q$ is the change in the number of the valence electron with UHV annealing, $\Delta$V$_M$ is the change in the Madelung potential, and $\Delta$E$_R$ is the change in the core-hole potential due to screening. In Fig. 4(a), we have plotted the binding energy shift of the O $1s$, Ti $2p$ and Sr $3d$ core-levels relative to 550$^\circ$C sample. The shifts in the binding energy of Au $4f_{7/2}$ core level were monitored throughout the experiment and plotted as a function of annealing temperature [Fig.4(a)]. The binding energy shift of O $1s$ and Sr $3d$ with annealing were in same directions while Ti $2p$ was different from them. Since the shifts of the core levels of the O$^{2-}$ anions and the Sr$^{2+}$ cations are in same directions the Madelung potential $\Delta$V$_M$ has negligible effects on the core-level shifts. The change in the number of the valence electron with UHV annealing ($K$$\Delta Q$) can often be excluded from the main origin of the core-level shifts since the Ti$^{3+}$ content is only $\sim$ 5$\%$ for 840 $^\circ$C sample. Therefore, we can assume that $\Delta$E$_B$ $\simeq$ $\Delta\mu$ + $\Delta$E$_R$. The O $1s$ and Sr $3d$ core levels shift toward lower binding energies with electron doping by creating O$_{vac}$. It would be expected from the chemical potential shift if the shifts of O $1s$ and Sr $3d$ core levels are in the opposite direction with electron doping. Therefore, we conclude that the shifts of the O $1s$ and Sr $3d$ core levels are largely due to the core-hole screening ($\Delta$E$_R$). The opposite shift of the Ti $2p$ core level can be explained by the the chemical potential shift. We have estimated $\Delta\mu$ using the Ti $2p$ core level shifted toward higher binding energies with annealing. In Fig. 4(b) we have plotted the chemical potential shift as a function of annealing temperature. We observe that the chemical potential monotonously moves upward with annealing temperature. Analysis of Ti $2p$ XPS core level spectra with UHV annealing reveals the presence of Ti$^{3+}$ component in addition to Ti$^{4+}$ in the metallic surface of SrTiO$_{3-\delta}$ while the Ti$^{3+}$ component are absent in the insulating SrTiO$_3$. According to electronic phase separation model the mixed valence of Ti ion in SrTiO$_{3-\delta}$ surface is comprised of coexisting nano-to-mesoscopic size cluster of metallic Ti$^{3+}$ rich and insulating Ti$^{4+}$ rich regions. By creating oxygen vacancies through annealing, electrons are doped, which naturally raise the chemical potential with increasing Ti$^{3+}$ component. In this scenario, if an electronic phase separation occurs then the chemical potential shift is most likely suppressed or lowered. On the other hand, the O $2p$ band in the valence band (Fig. 5) is not shifted, and therefore, the chemical potential is not pinned by the vacancy induced states formed within the band gap. The monotonous increase of chemical potential shift is consistent with the observation of progressive increasing of Ti$^{3+}$ component. Therefore, observed shift of the chemical potential with UHV annealing rules out the existence of electronic phase separation in SrTiO$_{3-\delta}$.

\begin{figure}[t]
\vskip 1.0cm
\includegraphics[width=4.0in]{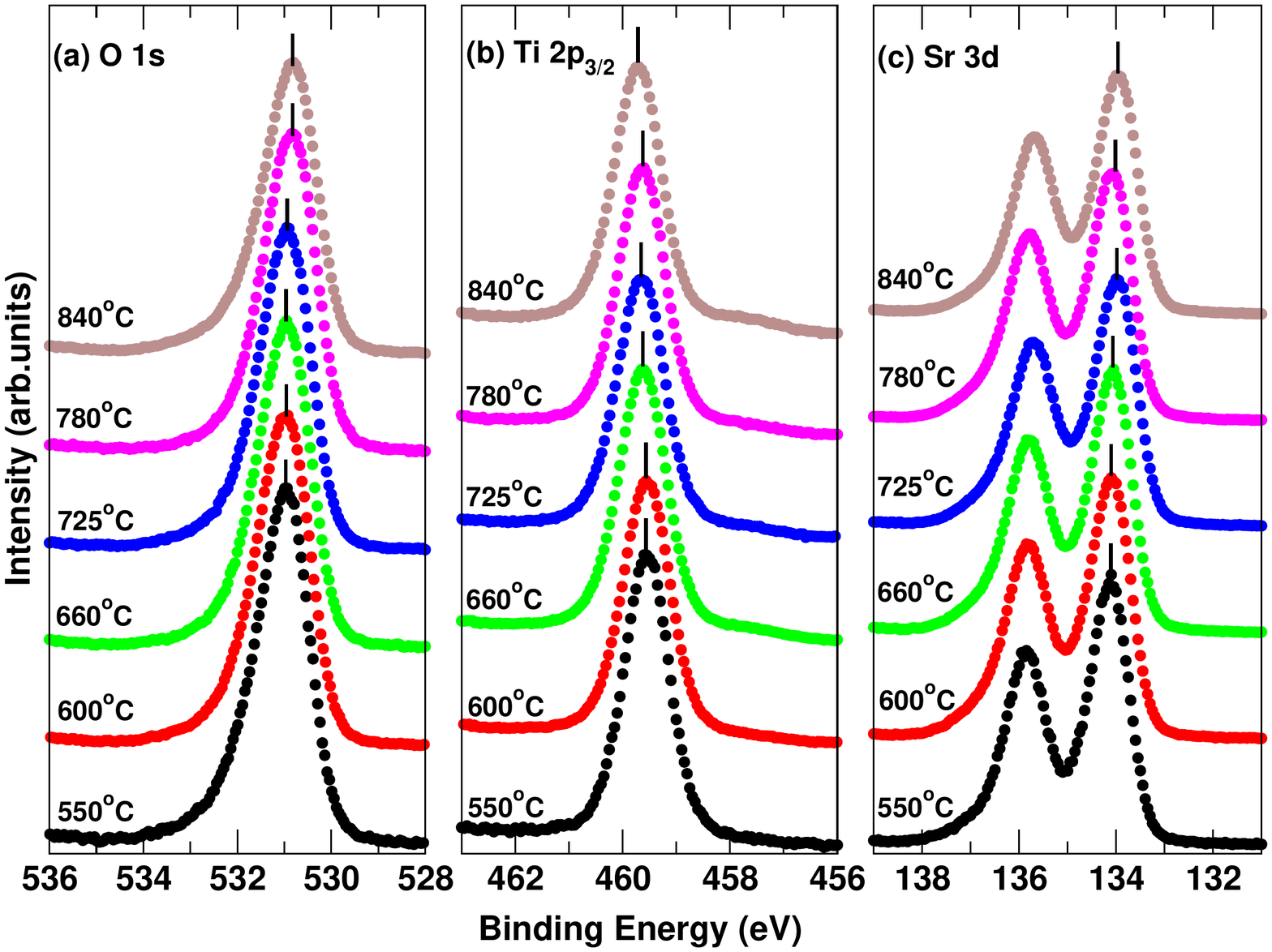}
\caption{\label{figure_3} (a) O $1s$ (b) Ti $2p_{3/2}$ and (c) Sr $3d$ core level photoemission spectra after different annealing temperatures taken using $1486.60$ eV photon energy. The O $1s$ and Sr $3d$ peak shifted to lower binding energy while Ti $2p$ shifted to higher binding energy with increasing annealing temperature.}
\end{figure}

Figure 5 shows the valence band (VB) photoemission spectra of the SrTiO$_3$ samples taken at different annealing temperatures. All the spectra are energy calibrated relative to the Fermi energy of Au film in electrical contact with the sample. Intensities of all the spectra were normalized and shifted along the ordinate axis by a constant value for the clarity of presentation. All the features seen in the spectra are dominated by the states due to Ti 3d - O 2p hybridization. The broad feature appearing at $\sim$ 7.1 eV (marked $C$) is due to the bonding states of this hybridization while its non-bonding states appear at $\sim$ 5.2 eV (marked $B$). A discussion on these spectral features can be found in earlier reported experiments and band structure calculations on similar systems \cite{fujimori01,ehre,kohiki}. At $1486.60$ eV photon energy, the emission features $B$ and $C$ are dominated by O $2p$ states due to their higher cross section compared to Ti $3d$ states \cite{lindau}. The figure also shows that there are no drastic changes in features $B$ and $C$ as a function of annealing, which indicates no electrons transfer among bonding and non-bonding states as annealing temperature is increased from 550 to 840$^\circ$C. The peak at $\sim$ 3.0 eV (marked $A'$) for 550, 600 and 660$^\circ$C samples is due to the mixture of Ti $3d$, $4s$ and $4p$ states of the distorted TiO$_6$ octahedra while these states for 725, 780 and 840$^\circ$C samples appear at $\sim$ 1.0 eV (marked $A$) \cite {lin1,lin2}. Above 725 $^\circ$C sample, the intensity of peak A near E$_F$ shows a substantial enhancement. The observed changes in these states will be discussed in the following paragraphs.

\begin{figure}[t]
\vskip 1.0cm
\includegraphics[width=4.0in]{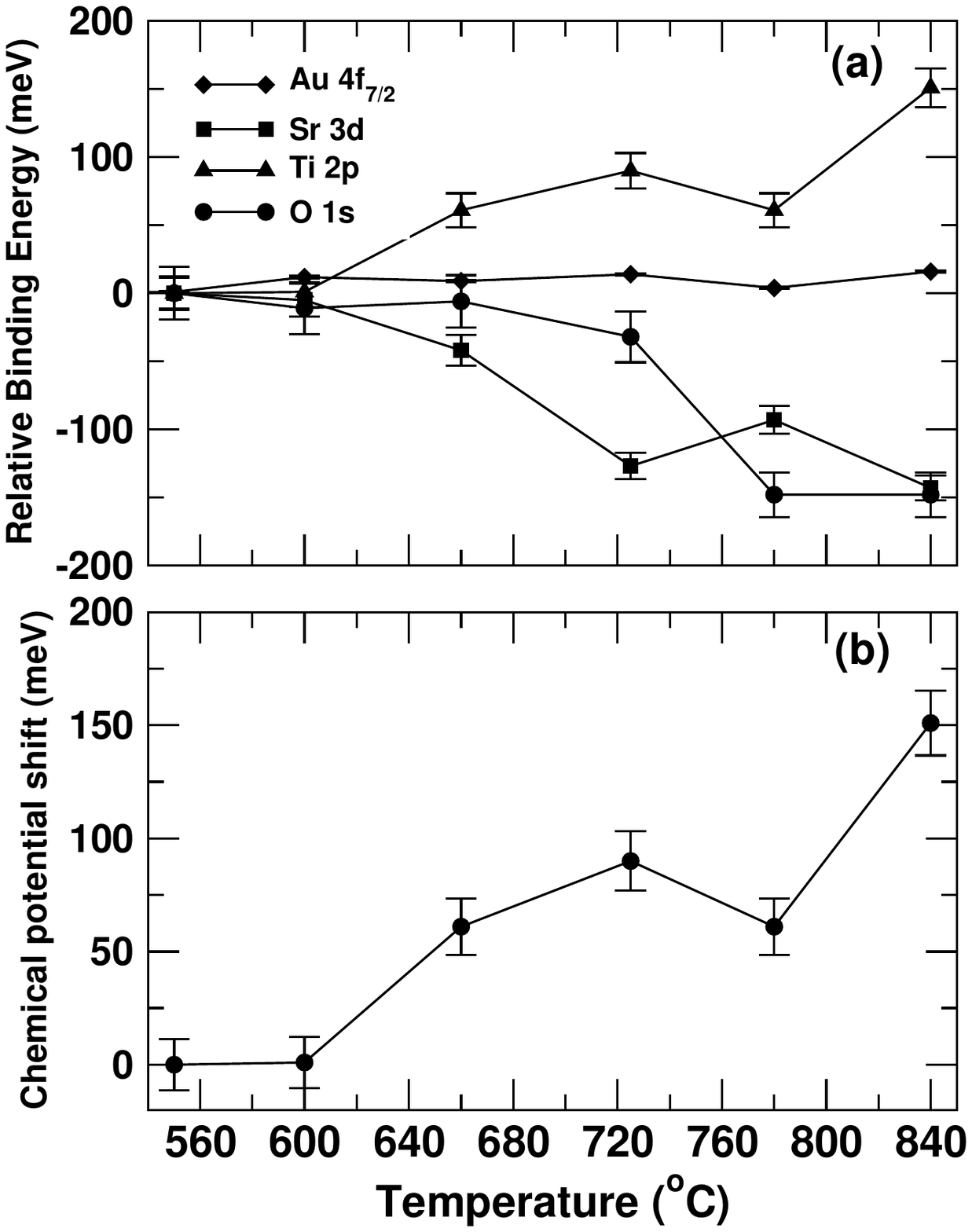}
\caption{\label{figure_4} (a) Binding energy shifts of the O $1s$, Ti $2p$ and Sr $3d$ core levels relative to 550$^\circ$C sample, which is thus set to zero, plotted as a function of annealing temperature. Au $4f_{7/2}$ core level shifts relative to 550$^\circ$C annealing temperature also plotted. (b) Chemical potential shift plotted as a function of annealing temperature. The shifts are negligible for samples heated between 550 to 600$^\circ$C.} 
\end{figure}

The expanded spectra of near E$_F$ region are shown in Fig. 6 (a) where annealing induced gap states are more clearly observed. Density of the gap states governs the materials physical properties. Here, we concentrate on the subtle changes in the near E$_F$ spectral features, marked as $A$ and $A'$. The 725, 780 and 840$^\circ$C samples have a significantly high spectral intensity at $A$ near the E$_F$, depicting the metallic nature of these samples. On the other hand, the 550, 600 and 660$^\circ$C samples display a soft gap at $A$. One can see that there is a shift in the spectral weight from $A'$ to $A$ as we increase the annealing temperature from 550 to 840$^\circ$C. Band structure calculations based on density functional theory (DFT) using the generalized gradient approximation (GGA) with a Hubbard {$\it U$} have shown that both these features arise from the Ti $3d$, $4s$ and $4p$ hybridized orbitals \cite{ishida,lin1,lin2,yoshida,higuchi2,aiura,koitzsch} due to the O$_{vac}$. For sake of comparison and clear manifest of the feature $A$, spectra corresponding to each annealing temperature plotted together with the 550 $^\circ$C sample, are shown in Fig. 6 (b). A close observation and comparison of Fig. 6(b) reveals that the spectra from 550, 600 and 660 $^\circ$C samples do not show any distinct feature corresponding to $A$ as in the case of the 725, 780 and 840$^\circ$C samples. The building up of DOS at $A$ is associated with the O$_{vac}$ in the SrTiO$_3$ system with progressive UHV annealing. The IM transition in SrTiO$_{3-\delta}$ was accompanied by the depletion of the feature at $A'$ due to the transfer of some DOS from $A'$ to $A$. When the annealing temperature is increased to 725 $^\circ$C the states melt to the position of $A$ due to the changes in the overlap of different Ti orbitals, as a consequence of structural changes via Ti$^{4+}$-O-Ti$^{3+}$ bond angles. The broadening of the feature $A$ becomes stronger with UHV annealing, resulting in increased delocalization of electronic states (metallic behavior) with increasing UHV annealing. The delocalization behavior of the charge carriers and thereby a shift of the spectral weight in the  near E$_F$ region, is generic to the IM transitions in electron doped SrTiO$_3$ with the presence of O$_{vac}$.

\begin{figure}[t]
\vskip 1.0cm
\includegraphics[width=4.0in]{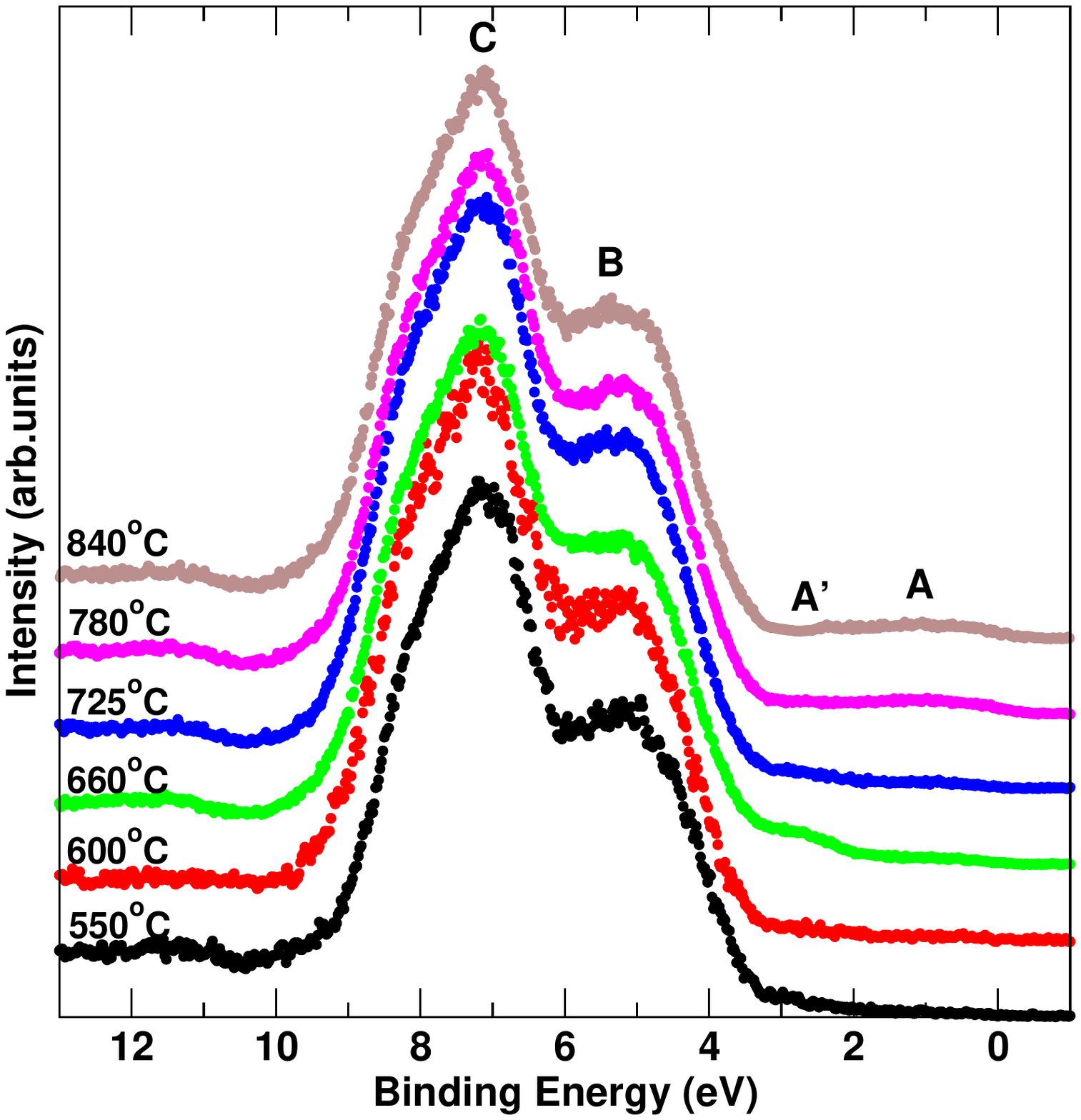}
\caption{\label{figure_5} The valence band photoemission spectra of the SrTiO$_3$(100) annealed at different temperatures under UHV conditions taken using monochoromatized Al K$\alpha$ line with photon energy $1486.60$ eV. The spectra are normalized and shifted along the y axis by a constant for clarity. The subbands around $1.0$, $3.0$, $5.2$ and $7.1$ eV are marked as $A$, $A'$, $B$ and $C$ respectively.}
\end{figure}

\begin{figure}[t]
\vskip 1.0cm
\includegraphics[width=4.0in]{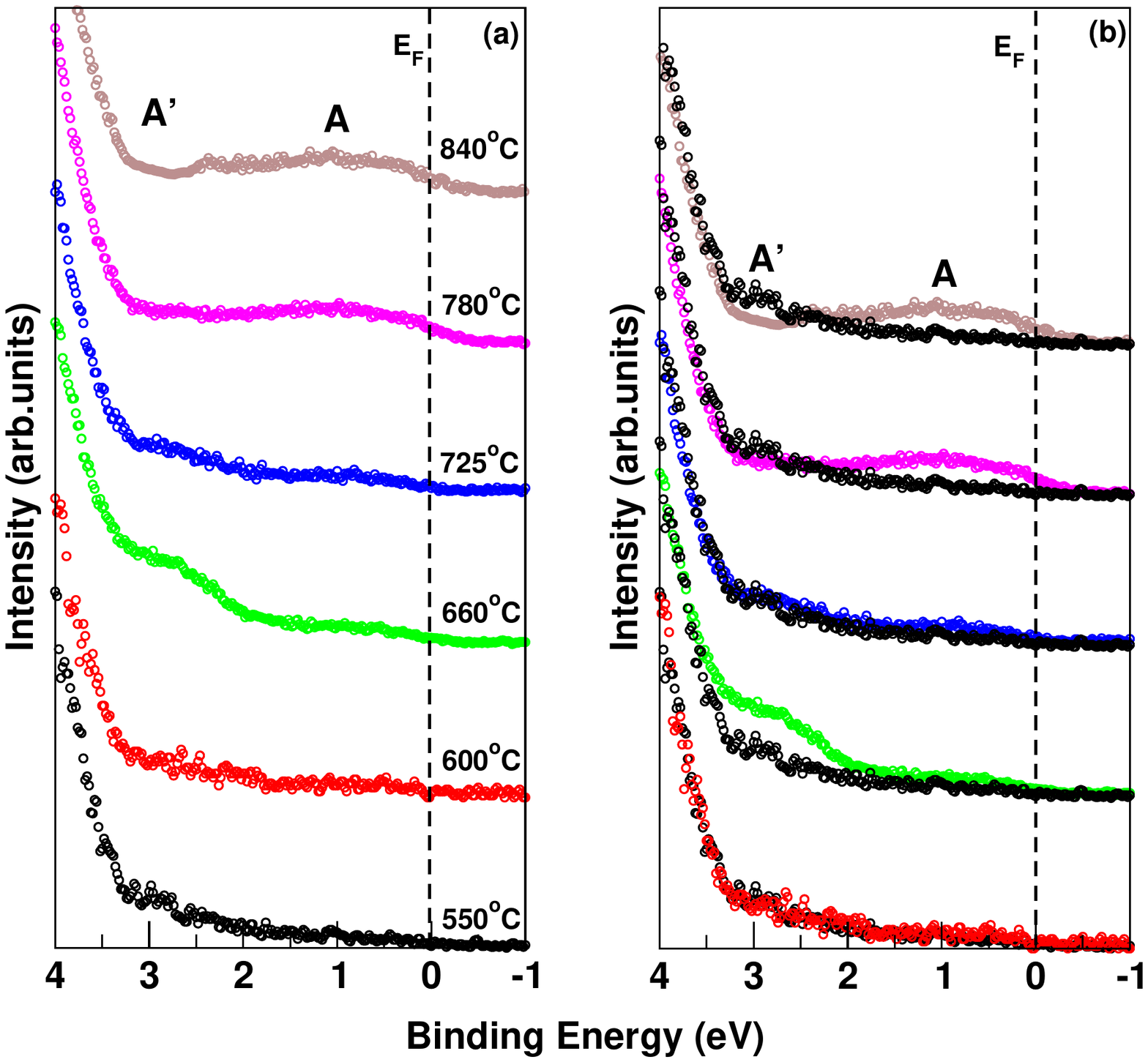}
\caption{\label{figure_6} (a) The near E$_F$ photoemission spectra presented as a function of annealing temperature. $A$ and $A'$ refer to the gap-states of the system. The 550, 600 and 660$^\circ$C samples have gap states $A'$ around $3.0$ eV while the 725, 780 and 840$^\circ$C samples have gap states $A$ around $1.0$ eV. The insulator-metal (IM) transition on increasing annealing temperature in this oxide is accompanied by the shifting of some DOS from from A' to A. (b) All of the other spectra plotted against the one from 550$^\circ$C sample, depicting the growth of $A$ with higher annealing temperatures.}
\end{figure}

The energy difference between the top of the valence band and the E$_F$ is about 3.2 eV which is approximately the band gap of the stoichiometric SrTiO$_3$. Therefore, the first available unoccupied states (conduction band edge) are located just below 0 eV binding energy. With an increase in annealing temperature the Ti$^{3+}$ content increases progressively and leads to a closing of the stoichiometric band gap. The IM transition in SrTiO$_{3-\delta}$ is due to the charge carriers of electrons of Ti$^{3+}$ state. The O$_{vac}$ create an extra charge near the TiO$_6$ octahedra with the creation of Ti$^{3+}$ component. The presence of Ti$^{3+}$ component can disrupt the TiO$_6$ octahedra by placing the Ti atoms in a slightly shifted position with respect to the oxygen cage and hence create a local electrical polarization \cite {lin2}. When Ti$^{3+}$ component is small, the extra electron may get trapped by the O$_{vac}$ and do not contribute towards conduction. In the temperature range between 550 - 660 $^\circ$C samples the O$_{vac}$ are small and are thus low carrier density (insulators). With its small O$_{vac}$ and accordingly small distortions, the charge carriers are localized. Consequently, this can result in the higher binding energies ($A'$) of the electronic states in the gap region. With an increase in Ti$^{3+}$ component the local electrical polarization increases and thus, the electrons become delocalized and contribute to the transport behavior of the samples. Therefore, in the temperature range between 725 - 840 $^\circ$C, those states are located in the near E$_F$ region ($A$). Thus, these states at $A$ for the 725, 780 and 840$^\circ$C samples appears broader suggesting that they are delocalized and contribute to the transport behavior of the samples. This reveals that annealing in the temperature range between 550 - 660 $^\circ$C results in gap-states right above the valence band maximum while high temperature annealing provides delocalized states in the near E$_F$ region.

\section{CONCLUSIONS}

We have investigated the electronic structure of SrTiO$_{3-\delta}$ using monochoromatized x-ray photoelectron spectroscopy. Annealing under UHV induces light electron doping via oxygen vacancies in stoichiometric SrTiO$_{3}$. Using core-level photoemission spectroscopy we have deduced the chemical potential shift as a function of annealing. With increase in annealing temperature the chemical potential monotonously moves upward which suggests that there is no electronic phase separation on SrTiO$_{3-\delta}$ and it rules out any sign of chemical potential pinning over the investigated temperature range. Further, we infer from our results as well as the results from the band structure calculations that the gap states have strongly mixed character of Ti $3d$, $4s$ and $4p$ states. A shift of DOS in the gap-states indicate the IM transition with electron doping as a function of UHV annealing. The building up of DOS as a function of annealing could be ascribed to the structural distortions induced by oxygen vacancies.

\section{ACKNOWLEDGEMENTS}

We thank Professor R. C. Budhani, Director NPL, India, for his kind support and constant encouragement. A.D. thanks IFCPAR-4704-1 and P. P. thanks CSIR Network Projects NWP-55 and PSC0109 for financial support.

\end{document}